# Agreeing on decisions: an analysis with counterfactuals

Bassel Tarbush
Department of Economics, University of Oxford
bassel.tarbush@economics.ox.ac.uk

## ABSTRACT

Moses & Nachum ([7]) identify conceptual flaws in Bacharach's generalization ([3]) of Aumann's seminal "agreeing to disagree" result ([1]). Essentially, Bacharach's framework requires agents' decision functions to be defined over events that are informationally meaningless for the agents. In this paper, we argue that the analysis of the agreement theorem should be carried out in information structures that can accommodate for counterfactual states. We therefore develop a method for constructing such "counterfactual structures" (starting from partitional structures), and prove a new agreement theorem within such structures. Furthermore, we show that our approach also resolves the conceptual flaws in the sense that, within our framework, decision functions are always only defined over events that are informationally meaningful for the agents.

## Categories and Subject Descriptors

J.4 [**Social and behavioral sciences**]: Economics; I.2.4 [**Knowledge Representation Formalisms and Methods**]: Frames and scripts

## General Terms

Theory

## Keywords

Agreeing to disagree, knowledge, belief, counterfactuals

## 1. INTRODUCTION

In [3], Bacharach generalized Aumann's seminal "agreeing to disagree" result ([1]) to the non-probabilistic case. Essentially, he isolated the relevant properties that hold of conditional probabilities, and of the common prior assumption - which drive the original result - and imposed them as independent conditions on general decision functions in partitional information structures. As such, he was able to isolate and interpret the underlying assumptions of the original result as (i) an assumption of "like-mindedness", which requires agents to take the same action given the same information, and (ii) an assumption that he claimed is analogous to requiring the agents' decision functions to satisfy Savage's Sure-Thing Principle ([9]). This principle is intended to capture the intuition that "if an agent takes the same action in every case when she is more informed, she takes the same action in the case when she is more ignorant".

However, in [7], Moses & Nachum found conceptual flaws in Bacharach's analysis, showing that his interpretations of "like-mindedness" and of the Sure-Thing Principle are problematic. Indeed, given that Bacharach is operating within partitional information structures, the information of agents is modeled as partitions of the state space. Furthermore, decision functions are defined over sets of states in a manner that is supposed to be consistent with the information that each agent has - in this way, decisions can be interpreted as being functions of agents' information. In Bacharach's set-up, like-mindedness requires the decision function of an agent $i$ to be defined over elements of the partitions of other agents $j$. But, except for the trivial case in which agent $i$'s partition element corresponds exactly to that of agent $j$, there is no sense in requiring $i$'s function to be defined over $j$'s partition element since that element is informationally meaningless to agent $i$. The Sure-Thing Principle is also problematic. An agent's decision function is said to satisfy the Sure-Thing Principle if whenever the decision over each element of a set of disjoint events is $x$, the decision over the union of all those events is also $x$. Notably, this implies that an agent's decision function must be defined over the union of her partition elements, but again, this is informationally meaningless for that agent since there is no partition element of that agent that corresponds to a union of her partition elements. More generally, Moses & Nachum show that Bacharach's set-up is such that the domains of the agents' decision functions contain elements that are informationally meaningless for the agents.

The basic premise of this paper is that the Sure-Thing Principle ought to be understood as an inherently counterfactual notion, and so any analysis that involves this principle but is carried out in an information structure that does not explicitly model the counterfactuals must be lacking in some way. Indeed, one could reformulate the intuition that the Sure-Thing Principle is intended to capture as: "If the agent takes the same action in every case when she is more informed, she *would* take the same action *if she were* more ignorant".

This distinction is important, but cannot be captured within Bacharach's framework because his analysis in [3] is carried out in partitional structures, and all information in those structures must be factual (in the sense that any belief that an agent holds must be true). In this paper, we therefore develop a method of transforming any given partitional structure into an information structure that explicitly includes the relevant counterfactual states. We interpret these "counterfactual structures" as being more complete pictures of the situation that is being modeled in the original partitional structure. The new set-up allows us to provide new formal definitions of the Sure-Thing Principle and of like-mindedness, that sit well with intuition, and we prove a new agreement theorem within these counterfactual structures.

Ultimately we show that our set-up resolves the conceptual issues raised by [7], in the sense that within counterfactual structures, decision functions are always only defined over events that are informationally meaningful for the agents.

In section 2 we present the formal definitions required to analyze information structures, and in section 3 we set up the framework of Bacharach, prove his version of the agreement theorem and provide Moses & Nachum's argument regarding the conceptual flaws. In section 4 we develop a method for constructing counterfactual structures, provide new definitions for the Sure-Thing Principle and for like-mindedness, and prove a new agreement theorem within such structures. Furthermore, we show that our approach resolves the conceptual flaws. Finally, in section 5 we relate our approach to other results and proposed solutions to the conceptual flaws found in the "agreeing to disagree" literature, and section 6 concludes. All proofs are in the appendix.

## 2. INFORMATION STRUCTURES

This section introduces the formal apparatus that will be used to derive the agreement theorem. In large part, the formal definitions given are completely standard.

### 2.1 General information structures

Let $\Omega$ denote a finite set of *states* and $N$ a finite set of agents. A subset $e \subseteq \Omega$ is called an *event*. For every agent $i \in N$, define a binary relation $R_i \subseteq \Omega \times \Omega$, called a *reachability* relation. So, we say that the state $\omega \in \Omega$ *reaches* the state $\omega' \in \Omega$ if $\omega R_i \omega'$. It terms of interpretation, if $\omega R_i \omega'$, then at $\omega$, agent $i$ considers the state $\omega'$ possible. An *information structure* $\mathcal{S} = (\Omega, N, \{R_i\}_{i \in N})$ is entirely determined by the state space, the set of agents, and the reachability relations.

The reachability relations $\{R_i\}_{i \in N}$ are said to be:

1. Serial if $\forall i \in N, \forall \omega \in \Omega, \exists \omega' \in \Omega, \omega R_i \omega'$.

2. Reflexive if $\forall i \in N, \forall \omega \in \Omega, \omega R_i \omega$.

3. Transitive if $\forall i \in N, \forall \omega, \omega', \omega''' \in \Omega$, if $\omega R_i \omega' \& \omega' R_i \omega''$, then $\omega R_i \omega''$.

4. Euclidean if $\forall i \in N, \forall \omega, \omega', \omega''' \in \Omega$, if $\omega R_i \omega' \& \omega R_i \omega''$, then $\omega' R_i \omega''$.

We have not yet imposed any particular restrictions on the reachability relations. We will therefore provide the definitions below in a general setting, with the understanding that they will only be applied in (i) **S5**, (ii) **KD45** and (iii) a special class of **KD4** structures. Respectively, this is when the reachability relations are (i) equivalence relations (reflexive and Euclidean), (ii) serial, transitive and Euclidean, and (iii) serial and transitive.

A *possibility set* at state $\omega$ for agent $i \in N$ is defined by

$$b_i(\omega) = \{\omega' \in \Omega | \omega R_i \omega'\} \tag{1}$$

A possibility set $b_i(\omega)$ is therefore, simply the set of all states that $i$ considers possible at $\omega$. In terms of notation, let us have $\mathcal{B}_i = \{b_i(\omega) | \omega \in \Omega\}$. For any $e \subseteq \Omega$, a *belief operator* is given by

$$B_i(e) = \{\omega \in \Omega | b_i(\omega) \subseteq e\} \tag{2}$$

Also, for any $e \subseteq \Omega$, and any $G \subseteq N$, a *mutual belief operator* is given by

$$M_G(e) = \cap_{i \in G} B_i(e) \tag{3}$$

This operator can be iterated by letting $M_G^1(e) = M_G(e)$ and $M_G^{m+1}(e) = M_G(M_G^m(e))$ for $m \geq 1$. For any $e \subseteq \Omega$, and any $G \subseteq N$, we can thus define a *common belief operator*,

$$C_G(e) = \cap_{m=1}^{\infty} M_G^m(e) \tag{4}$$

Finally, we say that a state $\omega' \in \Omega$ is *reachable* among the agents in $G$ from a state $\omega \in \Omega$ if there exists $\omega \equiv \omega_0, \omega_1, \omega_2, ..., \omega_n \equiv \omega'$ such that for each $k \in \{0, 1, ..., n-1\}$, there exists an agent $i \in G$ such that $\omega_k R_i \omega_{k+1}$. The *component* $T_G(\omega)$ (among the agents in $G$) of the state $\omega$ is the set of all states that are reachable among the agents in $G$ from $\omega$. Common belief can now be given an alternative characterization,

$$C_G(e) = \{\omega \in \Omega | T_G(\omega) \subseteq e\} \tag{5}$$

This is standard, and for example follows [6, p. 12].

### 2.2 Partitional structures

Consider an information structure $\mathcal{S} = (\Omega, N, \{R_i\}_{i \in N})$ and suppose that the reachability relations $\{R_i\}_{i \in N}$ are equivalence relations. Then, we say that $\mathcal{S}$ is a *partitional structure*. Indeed, the remark below shows that in this case, the information structure $\mathcal{S}$ becomes a standard partitional, or **S5**, or "knowledge" structure (see for example, [1]).

REMARK 1. *Suppose $\mathcal{S} = (\Omega, N, \{R_i\}_{i \in N})$ is a partitional structure. For any agent $i \in N$, $\omega \in b_i(\omega)$, and any $b_i(\omega)$ and $b_i(\omega')$ are either identical or disjoint; and, $\mathcal{B}_i$ is a partition of the state space.*

Note that in a partitional structure, at any state $\omega$, an agent $i$ considers any of the states in $b_i(\omega)$ (including $\omega$ itself) possible. The belief operator becomes the standard "knowledge" operator, and satisfies the following properties, which are well-known in the literature:

**K** $B_i(\neg e \cup f) \cap B_i(e) \subseteq B_i(f)$ \hfill *Kripke*
**D** $B_i(e) \subseteq \neg B_i(\neg e)$ \hfill *Consistency*
**T** $B_i(e) \subseteq e$ \hfill *Truth*
**4** $B_i(e) \subseteq B_i(B_i(e))$ \hfill *Positive Introspection*
**5** $\neg B_i(e) \subseteq B_i(\neg B_i(e))$ \hfill *Negative Introspection*

Note that in a partitional structure, the operator $C_G$ has the familiar interpretation of being the "common knowledge" operator. Furthermore, since this reduces to a completely standard framework, we easily obtain familiar results, such as the proposition below.

PROPOSITION 1. *Suppose $\mathcal{S} = (\Omega, N, \{R_i\}_{i \in N})$ is a partitional structure. Then, for any $\omega \in \Omega$ and any $i \in G$, $\cup_{\omega' \in T_G(\omega)} b_i(\omega') = T_G(\omega)$.*

## 2.3 Belief structures

Suppose now that the reachability relations $\{R_i\}_{i \in N}$ in an information structure $\mathcal{S} = (\Omega, N, \{R_i\}_{i \in N})$ are serial, transitive and Euclidean. Then, say we that $\mathcal{S}$ is a *belief structure*. Indeed, the information structure $\mathcal{S}$ becomes a standard **KD45** structure, for example, as presented in [6].

REMARK 2. *Suppose $\mathcal{S} = (\Omega, N, \{R_i\}_{i \in N})$ is a belief structure. For any agent $i \in N$, and any $\omega \in \Omega$, $b_i(\omega) \neq \emptyset$, and if $\omega \in b_i(\omega')$, then $b_i(\omega) = b_i(\omega')$.*

It is important to note that although every possibility set must be non-empty, it can be the case that $\omega \notin b_i(\omega)$. This means that at state $\omega$, agent $i$ considers states *other than $\omega$* to be possible, and not $\omega$ itself. The agent is therefore "deluded". (In fact, this terminology is directly borrowed from [6, p. 5]). Unsurprisingly, the belief operator now no longer satisfies the truth property **T**, but it does satisfy **K**, **D**, **4**, and **5**.

The salient point here is that the set-up presented has very close analogues in the literature, and allows us to drop - among other things - the property **T** of the belief operator, as compared with partitional structures. This will be important when including counterfactual states since by their very nature, these will be used to model information that can be false.

## 3. AGREEING ON DECISIONS

In this section, we present the original set-up of [3], derive his version of the agreement theorem, and then outline its inherent conceptual flaws which were originally raised in [7].

### 3.1 The original result

The original result was derived in a partitional information structure. The set-up in this entire section therefore assumes that we are working with a partitional structure $\mathcal{S} = (\Omega, N, \{R_i\}_{i \in N})$. Notably, this means that $\mathcal{B}_i$ is taken to be a partition of the state space for every agent $i \in N$ (see Remark 1).

For every agent $i \in N$, an *action function* $\delta_i : \Omega \to \mathcal{A}$, which maps from states to actions, specifies agent $i$'s action at any given state as a function of $i$'s possibility set at that state (which is intended to represent $i$'s "information" at that state); so the value of the action function will fully depend on the partition $\mathcal{B}_i$. A *decision function* $D_i$ for agent $i$, maps from a field $\mathcal{F}$ of subsets of $\Omega$ into a set $\mathcal{A}$ of actions. That is,

$$D_i : \mathcal{F} \to \mathcal{A} \qquad (6)$$

Following the terminology of [7], we will say that the agent $i$ using the action function $\delta_i$ *follows* the decision function $D_i$ if for all states $\omega \in \Omega$, $\delta_i(\omega) = D_i(b_i(\omega))$.

Bacharach imposes two main restrictions in order to derive his result, namely, the *Sure-Thing Principle* and *like-mindedness*. The definitions of these terms are given below.

*Definition 1.* The decision function $D_i$ of agent $i$ satisfies the *Sure-Thing Principle* if whenever for all $e \in \mathcal{E}$, $D_i(e) = x$ then $D_i(\cup_{e \in \mathcal{E}} e) = x$, where $\mathcal{E} \subseteq \mathcal{F}$ is a non-empty set of disjoint events.

In terms of interpretation, we can think of an event as representing some information and a decision over that event as determining the action that is taken as a function of that information. The union of events is intended to capture some form of "coarsening" of the information. So, following [7], the Sure-Thing Principle is intended to capture the intuition that "If the agent takes the same action in every case when she is more informed, she takes the same action in the case when she is more ignorant". Regarding like-mindedness, we have the following definition.

*Definition 2.* Agents are said to be *like-minded* if they have the same decision function.

That is, over the same subsets of states, the agents take the same action if they are like-minded. This is intended to capture the intuition that given the same information, the agents would take the same action.

THEOREM 1. *Let $\mathcal{S} = (\Omega, N, \{R_i\}_{i \in N})$ be a partitional structure. Then within $\mathcal{S}$, if the agents $i \in N$ are like-minded (as defined in Definition 2) and follow the decision functions $\{D_i\}_{i \in N}$ (as defined in (6)) that satisfy the Sure-Thing Principle (as defined in Definition 1), then for any $G \subseteq N$, if $C_G(\cap_{i \in G} \{\omega' \in \Omega | \delta_i(\omega') = x_i\}) \neq \emptyset$ then $x_i = x_j$ for all $i, j \in G$.*

This theorem states that if the actions taken by each member of a group of like-minded agents, who follow decision functions that satisfy the Sure-Thing Principle, are common knowledge among that group, then the members of the group must all take the same action. That is, the agents cannot "agree to disagree" about what action to take.

### 3.2 Conceptual flaws

[7] find conceptual flaws in the set-up of [3] outlined above. In broad terms, they find that the requirements that Bacharach imposes on the decision functions forces them to be defined over sets of states, the interpretation of which is meaningless within the information structure he is operating in. Formally, consider the following definition.

*Definition 3.* Let $\mathcal{S} = (\Omega, N, \{R_i\}_{i \in N})$ be some arbitrary information structure. We say that an event $e$ is a *possible belief* for agent $i$ in $\mathcal{S}$ if there exists a state $\omega \in \Omega$ such that $e = b_i(\omega)$.

When $\mathcal{S}$ is a partitional structure, this definition corresponds exactly to $e$ being a "possible state of knowledge" as defined in [7]. In [7], it is shown that

1. The Sure-Thing Principle forces decisions to be defined over unions of possibility sets, but no union of possibility sets can be a possible belief for any agent (see [7, Lemma 3.2]).

2. The assumption of like-mindedness forces the decision function of an agent $i$ to be defined over the possibility sets of agents $j \neq i$, but - other than the case when they correspond trivially - these are not possible beliefs for agent $i$ (see [7, Lemma 3.3]).

In other words, Bacharach's framework requires the decision functions to be defined over events that are not possible beliefs for the agents (within the information structure).

## 4. COUNTERFACTUAL STRUCTURES

The basic premise of this paper is that the Sure-Thing Principle ought to be understood as an inherently counterfactual notion, and so any analysis that involves this principle but is carried out in an information structure that does not explicitly model the counterfactuals must be lacking in some way. Indeed, one could reformulate the intuition that the Sure-Thing Principle is intended to capture as: "If the agent takes the same action in every case when she is more informed, she *would* take the same action *if she were* more ignorant" (where "more ignorant" has a well-defined meaning). This is counterfactual in the sense that there is no requirement for the agent to actually be more ignorant. Rather, the requirement is that the agent would take the same action in the situation where she imagines herself, counterfactually, to be more ignorant.

This distinction is important, but cannot be captured within Bacharach's framework. Indeed, the analysis in [3] is carried out in partitional structures. However, since the truth property **T** holds in such structures, every conceivable belief must be factual, and so by definition, counterfactual situations cannot be considered.[1] In this section, we therefore develop a method of transforming any given partitional structure into an information structure that explicitly includes the relevant counterfactual states. We interpret such "counterfactual structures" as being more complete pictures of the situation being modeled in the original partitional structure. We then provide new formal definitions for the Sure-Thing Principle and for like-mindedness and derive a new agreement theorem within these new structures. Ultimately this will resolve the conceptual issues raised by [7], in the sense that within counterfactual structures, decision functions are defined only over events that are possible beliefs for the agents.

### 4.1 Set-up with counterfactual states

In this section we define a method of transforming any given partitional structure into an information structure that explicitly includes the relevant counterfactual states.

It will be useful to introduce some new definitions. Suppose $\mathcal{S} = (\Omega, N, \{R_i\}_{i \in N})$ is a partitional structure. For every agent $i \in N$, define $I_i(\omega) = \{\omega' \in \Omega | \omega R_i \omega'\}$. Trivially, $I_i(\omega)$ is the equivalence class of the state $\omega$, and for each $i \in N$, $\mathcal{I}_i = \{I_i(\omega) | \omega \in \Omega\}$ is a partition of the state space (by Remark 1). Finally, let us define,

$$\Gamma_i = \{\cup_{e \in \mathcal{E}} e | \mathcal{E} \subseteq \mathcal{I}_i, \mathcal{E} \neq \emptyset\} \quad (7)$$

Clearly, $\Gamma_i$ consists of all the partition elements of $i$, and of all the possible unions across those partitions elements.

**Construction of counterfactuals.** Let $\mathcal{S} = (\Omega, N, \{R_i\}_{i \in N})$ be a partitional structure. We can immediately define $I_i(\omega) = \{\omega' \in \Omega | \omega R_i \omega'\}$, the partition $\mathcal{I}_i = \{I_i(\omega) | \omega \in \Omega\}$, and the set $\Gamma_i$ (described above) for every $i \in N$. From $\mathcal{S}$, we can create a new structure $\mathcal{S}' = (\Omega', N, \{R'_i\}_{i \in N})$, which we call the *counterfactual structure* of $\mathcal{S}$, where $\Omega' = \Omega \cup \Lambda$, $\Lambda$ is a set of states distinct from $\Omega$, and $R'_i \subseteq \Omega' \times \Omega$ is a reachability relation for every $i \in N$. The construction of the set $\Lambda$ and of the reachability relations $\{R'_i\}_{i \in N}$ is described below.

- For every $i \in N$, and for every $e \in \Gamma_i$, create a set $\Lambda_i^e$ of *new* states, which contains exactly one *duplicate* $\lambda_{i,\omega}^e$ of the state $\omega$ for every $\omega \in \Omega$ (so $|\Lambda_i^e| = |\Omega|$). We say that the *counterfactual state* $\lambda_{i,\omega}^e$ is the *counterfactual* of $\omega$ for agent $i$ with respect to the event $e$. The set of states $\Lambda$ is simply the set of all counterfactual states. Namely, $\Lambda = \cup_{i \in N} \cup_{e \in \Gamma_i} \Lambda_i^e$.[2]

- We now describe the process to construct the reachability relations $\{R'_i\}_{i \in N}$. For every agent $i \in N$, start with $R'_i = R_i$. We will add new elements to $R'_i$ according to the following method: For every $\lambda \in \Lambda$, if $\lambda = \lambda_{i,\omega}^e$ for some $\omega \in \Omega$ and $e \in \Gamma_i$, then (i) if $\omega \in e$ (that is, if $\lambda_{i,\omega}^e$ is the duplicate of a state in $e$), then for every $\omega' \in e$, add $(\lambda_{i,\omega}^e, \omega')$ as an element to $R'_i$, and (ii) if $\omega \notin e$, then for every $\omega' \in I_i(\omega)$, add $(\lambda_{i,\omega}^e, \omega')$ as an element to $R'_i$. Finally, if $\lambda = \lambda_{j,\omega}^e$ for some $\omega \in \Omega$, and $e \in \Gamma_j$ where $j \in N \setminus \{i\}$, then for every $\omega' \in I_i(\omega)$, add $(\lambda_{j,\omega}^e, \omega')$ as an element to $R'_i$. Nothing else is an element of $R'_i$.

This is best explained by means of an example. Consider a partitional structure $\mathcal{S}$ with $\Omega = \{\omega_0, \omega_1, \omega_2, \omega_3, \omega_4\}$, $N = \{a, b\}$, and partitions $\mathcal{I}_a$ and $\mathcal{I}_b$ as represented in Figure 1. In Figures 2-4, we represent a selection of substructures of the counterfactual structure $\mathcal{S}'$ of $\mathcal{S}$.[3] Figure 2 shows the set of counterfactual states $\Lambda_a^{\{\omega_3, \omega_4\}}$, as well as $\Omega$, and the reachability relations, $R'_i \subseteq \Lambda_a^{\{\omega_3, \omega_4\}} \times \Omega$, of both agents across these two sets. The reachability relations $R'_i \subseteq \Omega \times \Omega$ are left out, but they are unchanged (relative to $\mathcal{S}$) and therefore identical to what is shown in Figure 1. Note that each state in $\Lambda_a^{\{\omega_3, \omega_4\}}$ is simply a duplicate of a corresponding state in $\Omega$. For agent $b$, every state $\lambda_{a,\omega}^{\{\omega_3, \omega_4\}}$ simply points to all the states $\omega' \in I_b(\omega)$ (and nothing else). For agent $a$, every state $\lambda_{a,\omega}^{\{\omega_3, \omega_4\}}$ such that $\omega \in \{\omega_0, \omega_1, \omega_2\}$ simply points to all the states $\omega' \in I_a(\omega)$ (and nothing else). However, for a state $\omega \in \{\omega_3, \omega_4\}$, every state $\lambda_{a,\omega}^{\{\omega_3, \omega_4\}}$ points to both $\omega_3$ and $\omega_4$ (and nothing else), even though $I_i(\omega_3) \cap I_i(\omega_4) = \emptyset$. A similar patterns holds in Figures 3 and 4 which are there as additional examples for the reader. For practical reasons, we do not represent the full sets $\Lambda$ and $R'_i \subseteq \Omega' \times \Omega$ in a single diagram; and, note that even when taken together Figures 1-4 do not offer a complete picture of $\mathcal{S}'$.

The counterfactual structure of a partitional structure has several interesting properties, which we derive below.

PROPOSITION 2. *Suppose that $\mathcal{S}' = (\Omega', N, \{R'_i\}_{i \in N})$ is the counterfactual structure of a partitional structure $\mathcal{S} =*

---

[1] An agent $i$'s belief in an event $E$ if factual if $B_i(E) \subseteq E$.

[2] Note that the indexing of the sets $\Lambda_i^e$ by both $e$ and $i$ is crucial. Indeed, one must note that for any $i \in N$, and for any $e, e' \in \Gamma_i$ such that $e \neq e'$, $\Lambda_i^e \cap \Lambda_i^{e'} = \emptyset$. Furthermore, for any $i, j \in N$ such that $i \neq j$, if $e \in \Gamma_i$ and $e' \in \Gamma_j$, $\Lambda_i^e \cap \Lambda_j^{e'} = \emptyset$ (even if $e = e'$).

[3] Consider any two information structures $\mathcal{S}^+ = (\Omega^+, N, \{R_i^+\}_{i \in N})$ and $\mathcal{S}^- = (\Omega^-, N, \{R_i^-\}_{i \in N})$. We say that $\mathcal{S}^-$ is a *substructure* of $\mathcal{S}^+$ if $\Omega^- \subseteq \Omega^+$ and $R_i^- \subseteq R_i^+$ for every $i \in N$.

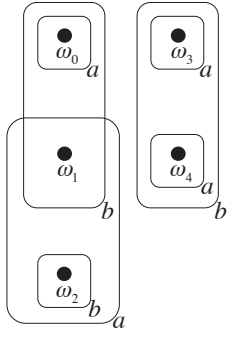

**Figure 1:** $\Omega$ and the partitions $\mathcal{I}_a$ and $\mathcal{I}_b$

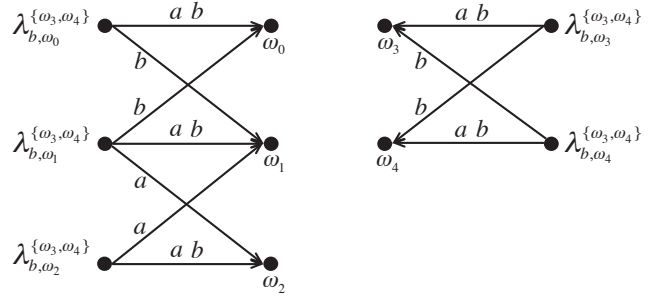

**Figure 3:** $\Lambda_b^{\{\omega_3,\omega_4\}} \cup \Omega$ and $R_i' \subseteq \Lambda_b^{\{\omega_3,\omega_4\}} \times \Omega$ for $i \in \{a,b\}$

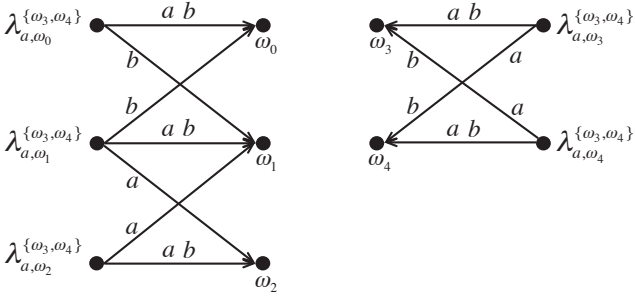

**Figure 2:** $\Lambda_a^{\{\omega_3,\omega_4\}} \cup \Omega$ and $R_i' \subseteq \Lambda_a^{\{\omega_3,\omega_4\}} \times \Omega$ for $i \in \{a,b\}$

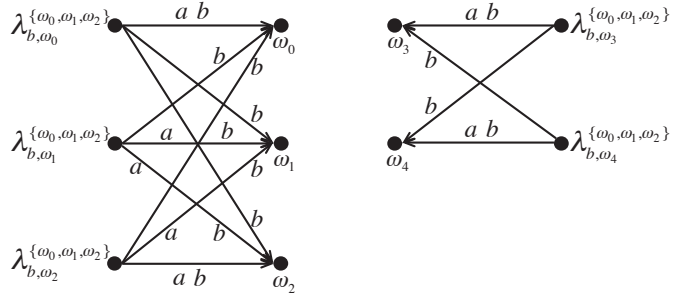

**Figure 4:** $\Lambda_b^{\{\omega_0,\omega_1,\omega_2\}} \cup \Omega$ and $R_i' \subseteq \Lambda_b^{\{\omega_0,\omega_1,\omega_2\}} \times \Omega$ for $i \in \{a,b\}$

$(\Omega, N, \{R_i\}_{i \in N})$. Then the reachability relations $\{R_i'\}_{i \in N}$ are serial and transitive.

PROPOSITION 3. Suppose that $\mathcal{S}' = (\Omega', N, \{R_i'\}_{i \in N})$ is the counterfactual structure of a partitional structure $\mathcal{S} = (\Omega, N, \{R_i\}_{i \in N})$. Then for any agent $i \in N$, (i) for any $\omega \in \Omega'$, $b_i(\omega) \neq \emptyset$, and if $\omega \in b_i(\omega')$, $b_i(\omega) \subseteq b_i(\omega')$, and (ii) for any $\omega \in \Omega$, $b_i(\omega) = I_i(\omega)$.

From the above, we have that counterfactual structures of partitional structures belong to the class of **KD4** structures. In particular, the belief operator now only satisfies properties **K**, **D**, and **4**; so "negative introspection" no longer holds, relative to belief structures. (See section 5.2 for further discussion of this point). Note however that within the counterfactual structure $\mathcal{S}' = (\Omega', N, \{R_i'\}_{i \in N})$ of a partitional structure $\mathcal{S} = (\Omega, N, \{R_i\}_{i \in N})$, the substructure $(\Omega, N, \{R_i\}_{i \in N})$ of $\mathcal{S}'$ corresponds exactly to the original structure $\mathcal{S}$ and is therefore partitional. A further result will be useful.

PROPOSITION 4. Suppose that $\mathcal{S}' = (\Omega', N, \{R_i'\}_{i \in N})$ is the counterfactual structure of a partitional structure $\mathcal{S} = (\Omega, N, \{R_i\}_{i \in N})$. Then for any $\omega \in \Omega'$ and any $G \subseteq N$, (i) if $\omega' \in T_G(\omega)$, then $\omega' \in \Omega$, and (ii) for any $i \in G$, $\cup_{\omega' \in T_G(\omega)} b_i(\omega') = T_G(\omega)$.

### 4.2 The agreement theorem

We will now adapt the main definitions required to derive the agreement theorem within the counterfactual structure of a partitional structure.

Throughout this section, we consider a partitional structure $\mathcal{S} = (\Omega, N, \{R_i\}_{i \in N})$, and the counterfactual structure $\mathcal{S}' = (\Omega', N, \{R_i'\}_{i \in N})$ of $\mathcal{S}$. As before, we can define $I_i(\omega) = \{\omega' \in \Omega | \omega R_i \omega'\}$, the partition $\mathcal{I}_i = \{I_i(\omega) | \omega \in \Omega\}$, and the set $\Gamma_i$ for every $i \in N$.

A decision function $D_i$ for an agent $i \in N$ maps from $\Gamma_i$ to a set of actions. That is,

$$D_i : \Gamma_i \to \mathcal{A} \qquad (8)$$

We now say that an action function $\delta_i : \Omega' \to \mathcal{A}$ *follows* decision function $D_i$ if for all states $\omega \in \Omega'$, $\delta_i(\omega) = D_i(b_i(\omega))$. The following proposition guarantees that this is well-defined.

PROPOSITION 5. Suppose that $\mathcal{S}' = (\Omega', N, \{R_i'\}_{i \in N})$ is the counterfactual structure of a partitional structure $\mathcal{S} = (\Omega, N, \{R_i\}_{i \in N})$. Then for any $\omega \in \Omega'$, $b_i(\omega) \in \Gamma_i$.

Below, we provide definitions for the Sure-Thing Principle and like-mindedness that are analogous to the ones proposed by Bacharach. We elaborate on their interpretations in section 4.4.

*Definition 4.* The decision function $D_i$ of agent $i$ satisfies the *Sure-Thing Principle* if for any non-empty subset $\mathcal{E}$ of $\mathcal{I}_i$, whenever for all $e \in \mathcal{E}$, $D_i(e) = x$ then $D_i(\cup_{e \in \mathcal{E}} e) = x$.

The domain $\Gamma_i$ includes all possible unions of elements of the partition $\mathcal{I}_i$, so this is well-defined. Furthermore, note that $\mathcal{E}$ must be a set of disjoint events.[4]

*Definition 5.* Agents $i$ and $j$ are said to be *like-minded* if for any $e \in \Gamma_i$ and any $e' \in \Gamma_j$, if $e = e'$ then $D_i(e) =$

---

[4]This contrasts with [7] who, in their solution, propose adopting a version of the Sure-Thing Principle that does not require the disjointness of events.

$D_j(e')$.[5]

THEOREM 2. *Let $\mathcal{S}' = (\Omega', N, \{R'_i\}_{i \in N})$ be the counterfactual structure of a partitional structure $\mathcal{S} = (\Omega, N, \{R_i\}_{i \in N})$. Then, within $\mathcal{S}'$, if the agents $i \in N$ are like-minded (as defined in Definition 5) and follow the decision functions $\{D_i\}_{i \in N}$ (as defined in (8)) that satisfy the Sure-Thing Principle (as defined in Definition 4), then for any $G \subseteq N$, if $C_G(\cap_{i \in G}\{\omega' \in \Omega' | \delta_i(\omega') = x_i\}) \neq \emptyset$ then $x_i = x_j$ for all $i, j \in G$.*

Although this agreement theorem might appear to have many similarities with the previous one, it is conceptually entirely distinct. In particular, we show below (in section 4.3) that we were able to obtain the result while avoiding the conceptual flaws that were discussed in section 3.2. We also provide an interpretation of Theorem 2 and of counterfactual structures of partitional structures more generally in section 4.4.

### 4.3 Solution to the conceptual flaws

As discussed in section 3.2, Bacharach's framework requires the decision functions to be defined over events that are not possible beliefs for the agents. The proposition below shows that this is not the case in our set-up.

PROPOSITION 6. *Suppose that $\mathcal{S}' = (\Omega', N, \{R'_i\}_{i \in N})$ is the counterfactual structure of a partitional structure $\mathcal{S} = (\Omega, N, \{R_i\}_{i \in N})$. Then for any $e \in \Gamma_i$, there exists an $\omega \in \Omega'$ such that $b_i(\omega) = e$. (In fact, there exists a state $\lambda^e_{i,\omega} \in \Lambda$ for some $\omega \in e$ such that $b_i(\lambda^e_{i,\omega}) = e$).*

This proposition, in conjunction with Proposition 5, shows that in our set-up, the domain of the decision function of every agent is exactly the set of all possible beliefs for that agent. Indeed, our decision functions are defined over unions of partition elements, but these are possible beliefs for the agents because for every such union, there exists a counterfactual state at which the possibility set is precisely that union. We therefore avoid the first point in the conceptual flaws raised by [7]. Regarding the second point, the decision function $D_i$ of agent $i$ is now only defined over events in $\Gamma_i$. There is therefore no requirement for the function to determine the agent's action in the case where the event corresponds to a partition element of another agent.

### 4.4 Interpretation

In this section, we provide an interpretation of our assumptions, showing that the formal definitions of the Sure-Thing Principle and of like-mindedness given in our set-up match well with intuition. We also provide an interpretation of the agreement theorem in counterfactual structures, and of those structures more generally.

Our notion of like-mindedness is straightforward: Over the same information, like-minded agents take the same action. However, our definition has an advantage over Bacharach's which is that an agent $i$ is not required to consider what action to take over the partition elements of another agent $j$.

With regards to the Sure-Thing Principle, the proposition below, in particular part (ii), allows us to interpret our version of the principle as capturing the intuition that: "If the agent takes the same action in every case when she is more informed, she *would* take the same action *if she were* (secretly) more ignorant".

PROPOSITION 7. *Suppose that $\mathcal{S}' = (\Omega', N, \{R'_i\}_{i \in N})$ is the counterfactual structure of a partitional structure $\mathcal{S} = (\Omega, N, \{R_i\}_{i \in N})$. Then, (i) for any $e \subseteq \Omega'$, and $\omega, \omega' \in \Omega'$, $b_i(\omega) \subseteq e$ and $b_i(\omega') \subseteq e$ if and only if $b_i(\lambda^{b_i(\omega) \cup b_i(\omega')}_{i,\omega''}) \subseteq e$ (for some $\omega'' \in \Omega$). (ii) For any $e \subseteq \Omega'$, and $\omega, \omega' \in \Omega$, $b_i(\omega) \subseteq e$ and $b_i(\omega') \subseteq e$ if and only if $b_i(\lambda^{b_i(\omega) \cup b_i(\omega')}_{i,\omega}) \subseteq e$.*

Indeed, suppose $\mathcal{S}' = (\Omega', N, \{R'_i\}_{i \in N})$ is the counterfactual structure of a partitional structure $\mathcal{S} = (\Omega, N, \{R_i\}_{i \in N})$. Now consider an agent $i$, and two partition elements $I_i(\omega)$, $I_i(\omega') \in \mathcal{I}_i$ (where $\omega, \omega' \in \Omega$), and suppose that her decision function is such that $D_i(I_i(\omega)) = D_i(I_i(\omega')) = x$. The Sure-Thing Principle requires that $D_i(I_i(\omega) \cup I_i(\omega')) = x$. Propositions 6 shows that the possibility set that corresponds to $I_i(\omega) \cup I_i(\omega')$ is $b_i(\lambda^{I_i(\omega) \cup I_i(\omega')}_{i,\omega})$. Proposition 7 part (ii) shows that for any event $e$, $i$ believes $e$ at the counterfactual state $\lambda^{I_i(\omega) \cup I_i(\omega')}_{i,\omega}$ if and only if $i$ also believes $e$ at the states within *each* of those partition elements. Informally, if we can call a belief in an event "information", then the information that $i$ has at the counterfactual state preserves only the information that is the same across both the partition elements. In this sense, the information that $i$ has at the counterfactual state is the information that $i$ would have if she were *just* more ignorant than at a state in either of the partition elements.[6] Furthermore, by construction of counterfactual structures, there is no state $\omega''' \in \Omega'$ and no $j \in N$ such that $(\omega''', \lambda^{I_i(\omega) \cup I_i(\omega')}_{i,\omega}) \in R'_j$; and, for any $j \neq i$, $(\lambda^{I_i(\omega) \cup I_i(\omega')}_{i,\omega}, \omega''') \in R'_j$ for every $\omega''' \in I_j(\omega)$ *only*. In words, this means that at this counterfactual state, $i$ may have become "more ignorant", but the information of all other agents is unchanged. The information at this state therefore truly captures the fact that $i$ is imagining herself *secretly* to be more ignorant. The situation is counterfactual since all other agents still believe that $i$ has the information that she does in the partition $\mathcal{I}_i$.

We believe that this interpretation of the Sure-Thing Principle matches well with intuition. In particular, given that the principle finds its origins in single-agent decision theory (see [9]), it makes sense that the requirement on the decisions in cases where the agents are more ignorant is imposed only when ignorance is secret - in the sense that the information of all other agents is unchanged.

More generally, our interpretation of the counterfactual structure $\mathcal{S}'$ of a partitional structure $\mathcal{S}$ is therefore that it is simply a more complete picture of the situation that is being modeled by the structure $\mathcal{S}$ since it also includes states in which the agents imagine themselves (secretly, and counterfactually) to be more ignorant. The inclusion of these states turns out to be relevant in deriving appropriate formal definitions of the Sure-Thing Principle and of like-mindedness, and in resolving the conceptual flaws regarding the domain of the decision functions. Indeed, we can think of the substructure $\mathcal{S}$ of $\mathcal{S}'$ as representing the "actual" situation, and the counterfactual states $\Lambda$ are essentially "fake" in the sense that they do not actually occur. However, they are connected to the "actual" states in $\Omega$ in a manner that captures every possible way in which every agent could be secretly more ignorant relative to the "actual" situation; and although the "fake" states do not occur, the decision functions are essentially defined at such states (More precisely, they are defined over possibility sets that are defined as such states).[7] This turns out to be crucial: Theorem 2 is derived by showing that when the actions of agents are commonly known, the Sure-Thing Principle and like-mindedness imply that the actions must be the same precisely in the case when the decision functions are based on the information at some counterfactual (or "fake") states. The equality at the counterfactual states then carries over to the decisions over the information in the "actual" situation, and therefore agents cannot agree to disagree.

---

[5]In contrast with the previous definition, we do not say that agents are like-minded if they have the "same" decision functions since the domains of the decision functions will now typically be different for different agents.

[6]In fact, it corresponds to being *just* "less informed", in a sense similar to that given in [8].

# 5. RELATION TO THE LITERATURE

We now discuss our approach in relation to other solutions that were proposed regarding the conceptual flaws. We then also compare our construction of the counterfactual states to other models that carry out a related exercise.

## 5.1 Other solutions

[7] propose a solution to the conceptual flaws that they found in the result of [3]. Essentially, they define a "relevance projection", which maps from sets of states to the "relevant information" at that set of states (see [7, p. 158]). They then impose conditions on this projection and on the decision functions to derive a new agreement theorem. However, it is not always obvious how a projection satisfying their conditions ought to be found. In contrast, the approach presented here offers a *constructive method* of obtaining a structure in which the analysis can be carried out.[8] Furthermore, our Sure-Thing Principle does require the disjointness of events, which their version does not.

[2] also propose a solution using a purely syntactic approach. The approach presented here is completely set-theoretic. Furthermore, they impose the condition that higher-order information must be irrelevant to the agents' decision, which we do not impose here.

Finally, [8] presented a very interesting solution to the conceptual flaws by redefining the Sure-Thing Principle entirely. Roughly, Samet's "Interpersonal Sure-Thing Principle" states that if agent $i$ knows that agent $j$ is more informed than he is, and knows that $j$'s action is $x$, then $i$ takes action $x$. Combining this with the assumption of the existence of an "epistemic dummy" - an agent who is less informed than every other agent - [8] proves a new agreement theorem in paritional structures. The large differences in the assumptions make a formal comparison between the approach here and in [8] difficult.

## 5.2 Action models

Loosely speaking, it was shown that the information at the counterfactual states in a counterfactual structure corresponds to secretly "losing" information. It turns out that secretly "gaining" information is well-studied in the dynamic epistemic logic literature (e.g. [4]). *Action models* formalize how the underlying structure (both the state space and the reachability relations) must be modified to model various protocols by which agents may gain some new information.

It was shown, [11, Theorem 17], that in the case of secretly gaining new information, a partitional structure would have to be transformed into a belief structure. In this paper, we have defined a method of modeling secret loss of information by transforming a partitional structure into a (counterfactual) structure that belongs to the **KD4** class. In particular, this means that "negative introspection" is dropped as a property of the belief operator. We have not shown that it is necessary to drop negative introspection in order to model secret loss of information, so in principle, it remains an open question as to whether it is possible to define a purely semantic transformation of a partitional structure (i.e. only involving the states and the reachability relations) that can model secret loss of information where the resulting structure is a belief structure in which the primitives of the original model (i.e. the original state space and partitions over them) are unchanged.[9]

## 5.3 Counterfactuals

General set-theoretic information structures have been proposed to model counterfactuals (e.g. see [5]), especially in relation to the literature on backwards induction. In extensive form games, to implement the backwards induction solution, agents must consider what they would do at histories of the game that might never be reached. They must therefore be able to define what they would do in situations that never occur. This therefore bears some resemblance to our set-up in which agents are required to have decisions that are defined over information at counterfactual (or "fake") states that never actually occur, but there are important differences which we briefly outline below.

There is a multitude of ways in which counterfactuals can be modeled, and we cannot hope to survey the literature here. However, it will suffice to say that a general approach to modeling counterfactuals proceeds in roughly the following manner: One defines a "closeness" relation on states and then says that a state $\omega$ belongs to the event "If $f$ were the case, then $e$ would be true" if $e$ is true in all the closest states to $\omega$ where $f$ is true. It is possible to then augment this approach with epistemic operators and decisions, but the salient point is simply that the standard approach to counterfactuals aims to be quite general, in capturing all possible hypothetical situations $f$.

In contrast, we only model counterfactuals for a very particular set of hypothetical situations, namely, every possible situation (relative to the "actual" situation) in which every agent considers herself to be secretly more ignorant. This is not done by imposing a closeness relation, but by creating a new set of "fake" counterfactual states and carefully re-wiring them to the "actual" states. (Note however, that the resulting information at the counterfactual states was shown to be interpretable as being secretly "just" more ignorant than in the "actual" situation being considered, so in this sense, the counterfactual state can be seen as being "close" to the actual situation). As a result, it is not obvious to see how the method developed here can be applied to studying backwards induction, which requires considering a richer set of hypothetical situations, but the method is well-adapted for the analysis of agreement theorems carried out in this paper.

---

[7]Notice that this shows that our counterfactual structures are particular "impossible-world" structures (e.g. see [12]). We return to this point in section 5.

[8]Also, the resulting counterfactual structure does satisfy properties that resemble, in spirit, the conditions imposed on the relevance projection.

[9][10] analyzes counterfactuals in **KD45** structures. However, his *initial* structures are **KD45**, whereas the point made here is regarding a method that would transform a *partitional* structure into a **KD45** structure while building the relevant counterfactual states and leaving the primitives of the original model unchanged.

Note that there is another approach to modeling counterfactuals that is related to ours. What is known as the "impossible-worlds" approach (e.g. [12]) augments information structures with a new set of states and with modified reachability relations. The set of states in the original structure are then referred to as "possible", or "normal", worlds, while the ones in the new set are referred to as "impossible", or "non-normal". In our framework, these actually correspond to our "actual" states $\Omega$, and our "fake" states $\Lambda$ (and the reachability relations are modified from $R_i$ to $R'_i$ for every $i$). The counterfactual structures presented here can therefore be seen as specific "impossible-worlds" structures. However, we are not aware of any paper that use impossible-worlds structures as a tool for modeling counterfactuals in the manner presented here.

## 6. CONCLUSION

We provided a *constructive method* for creating an information structure that includes the relevant counterfactual states (starting from a partitional structure). This new *counterfactual structure* is interpreted as providing a more complete picture of the situation that is being modeled by the original partitional structure. As such, our analysis of the agreement theorem is carried out in such structures.

Having provided new formal definitions for the Sure-Thing Principle and for like-mindedness, we prove an agreement theorem within such structures, and show that we can interpret our version of the Sure-Thing principle as capturing the intuition that: "If the agent takes the same action in every case when she is more informed, she *would* take the same action *if she were* (*secretly*) more ignorant". We also show that our version of like-mindedness has more desirable properties than Bacharach's. Furthermore, we show that our approach resolves the conceptual issues raised by [7], in the sense that within counterfactual structures, decision functions are defined only over events that are possible beliefs for the agents.

Therefore, in providing a constructive method for creating counterfactual structures, our approach achieves the goal of maintaining an interpretation of the underlying assumptions of the agreement theorem that fits well with intuition, while simultaneously resolving the conceptual issues (identified in [7]) regarding the domain of the decision functions.

## 7. ACKNOWLEDGMENTS

I would like to thank John Quah for his invaluable help, Francis Dennig and Alex Teytelboym for very useful discussions, as well as Harvey Lederman and three anonymous referees for their comments. I would also like to thank Dov Samet for conversations on earlier versions of the ideas presented here, and Burkhard Schipper for his helpful comments.